\documentclass[11pt,twoside]{article}
\usepackage{asp2004}
\usepackage{psfig}
\usepackage{epsf}
\usepackage{graphics}
\usepackage{lscape}
\def\lesssim{\mathrel{\hbox{\rlap{\hbox{\lower4pt\hbox{$\sim$}}}\hbox{$<$}}}}
\def\gtrsim{\mathrel{\hbox{\rlap{\hbox{\lower4pt\hbox{$\sim$}}}\hbox{$>$}}}}

\begin{document}

\title{Percolation Properties of Nearby Large-Scale Structures: Every Galaxy has a Neighbour}

\author{A.P. {Fairall}, D. {Turner}, M.L. {Pretorius}, M. {Wiehahn}, V. {McBride}, G. {de Vaux} and P.A. {Woudt}} 
\affil{Department of Astronomy, University of Cape Town, Private Bag,
        Rondebosch 7700, South Africa}

\begin{abstract}
The distribution of nearby (cz $<$ 7500 km s$^{-1}$) galaxies has been explored by minimal spanning trees; allowance has been made for the drastic decrease of data with distance.
The investigation finds that all galaxies are members of irregular elongated large-scale structures; there are no `field galaxies'. Based on our local large-scale structure, every galaxy appears to 
have a neighbouring galaxy within $<$100 km s$^{-1}$ (1.4 $h_{70}^{-1}$ Mpc) of redshift space, and thereby all galaxies are found to lie in filaments or tree configurations.

Every large-scale structure appears to have a neighbouring large-scale structure within $<$700 km 
s$^{-1}$ (10 $h_{70}^{-1}$ Mpc), such that large-scale structures interconnect to form a continuous labyrinth.
\end{abstract}

\section{Introduction}

The texture displayed by the Universe today is one of a labyrinth of large-scale structures and 
voids - first described as `cellular' by Joeveer and Einasto (1978). It is this labyrinth of 
large-scale structures and voids -- the ``Cosmic web'' -- that is the focus of much of this meeting (and appropriate that 
pioneers such as Jaan Einasto and John Huchra are present).

From mappings of local large-scale structures (e.g. Fairall 1998, hereafter F98), we now know that 
the {voids} range in size up to 85 $h_{70}^{-1}$ Mpc (6000 km s$^{-1}$ in redshift space); the closest such large void 
being the well-known {Sculptor Void} (R.A. 0.5h, Decl. -35deg, $cz = 5500$ km s$^{-1}$). Voids have a tendency 
towards sphericity; they have often been described as bubble-like. Like the bubbles in a bath sponge, 
they also interconnect (on the basis that interconnecting holes are larger than the correlation length 
of the galaxies surrounding the voids -- e.g. Gott et al.~1986). Were it possible, one could 
travel throughout the universe, passing from void to void. The most clearly defined voids 
(e.g. Sculptor, Microscopium) are empty, or have substantial empty cores. Other voids look 
as if they were formed from mergers of smaller voids, with the remnants of previous boundaries still apparent. 

However, within the research community, there does not seem to be consensus as to how voids 
should be defined. Some researchers see them as truly `void', others as merely under-densities. 
Our choice of a working definition for voids is simple: they are empty (Fairall et al.~1991). 
As the Local Void so clearly demonstrates, there is no population of low luminosity galaxies 
filling the voids. In the past, we have used a void-finding algorithm that fitted empty, approximately 
spherical, progressively smaller, voids into the distribution of galaxies (Kauffmann \& Fairall 1991). 
Where voids appear to have resulted from mergers, then multiple empty voids are fitted to fill the cavity.

Large-scale structures are gigantic conglomerations of galaxies. They have irregular shapes, 
since they border on the bubbly voids, or even incorporate small voids or holes. However there 
is a tendency to be flattened (walls) or ribbon-like. Large-scale structures interconnect 
with one another to form the labyrinth.

The galaxies within the large-scale structures tend to lie in chains or filaments. They wrap around the voids in a fashion similar to the filaments that wrap around the hollow interior of the Crab Nebula. It is this tendency that has given rise to the texture being described as the `Cosmic Web' (e.g. Lahav, Karachentsev, these proceedings).

Is there a way to define a large-scale structure? The first author (F98) has proposed a working 
definition that large-scale structures are formed by conglomerations of galaxies, where 
galaxy-galaxy separations are less that 200 km s$^{-1}$ ($\sim$ 3 $h_{70}^{-1}$ Mpc). However, we shall show below that, 
once a reasonably complete sample is used, the separations may even be lower than 100 km s$^{-1}$ (1.4 $h_{70}^{-1}$ Mpc).

The definition is based on the idea that every galaxy has a neighbour, the impression gained by the 
first author (when preparing the maps for the {\em Atlas of Nearby Large-Scale Structures} - F98), 
particularly for very nearby galaxies, where data were most complete. A further impression gained 
was that there might not be isolated islands of galaxies. Similarly, all large-scale structures 
might interconnect and none be isolated. If so, all the high density regions, in the local Universe 
at least, would be linked, and one could travel around in the universe by keeping to high density 
regions (again much like the material that makes a bath sponge). In Section 3, we attempt to 
quantify this situation by means of Minimal Spanning Trees.

\section{Data and incompleteness}

Given the enormous success of the 2dF and Sloan Digital Sky Surveys, and their predecessors, 
many researchers might consider their databases appropriate for investigating the characteristics 
of large-scale structures. What is not often appreciated is that such traditional (magnitude-limited) redshift 
surveys sample at best only the brightest few galaxies of several hundred, and possibly only 
the brightest few galaxies of several thousand. This will be apparent when we examine local data below.

Ideally, what we need to test the `neighbour' hypothesis is a {volume-limited sample} -- i.e. a 
sample volume of space where all galaxies that exist within that volume are known. Unfortunately 
there is no such thing as a volume-limited sample, short perhaps 
of our local group of galaxies, and even there, new members are still discovered 
from time to time (such as the recent finding of Andromeda VIII, Morrison et al.~2003). 
The {\em Catalogue of Nearby Galaxies}, announced at this meeting by Karachentsev, is probably 
the closest approximation we have to a volume-limited sample. Unfortunately, it comes too 
late for the present investigation, but will be used in a follow up.

What is the closest approximation to a volume-limited sample for nearby large-scale 
structures? The answer is to use all-available redshifts. For this reason we have turned 
to the NASA/IPAC Extragalactic Database (NED). Critics may argue that this is a statistically 
uncontrolled sample. Admittedly so, but all catalogues suffer at various levels from incompleteness (some are highly incomplete). In this light, the best one can do is to use the 
highest number of redshifts available. Furthermore, by restricting our investigation 
to very shallow redshifts, there is good all-sky coverage, without major selection effects in direction.

The data used below contains all available redshifts with $cz \le 7500$ km s$^{-1}$, as extracted 
from NED in April 2003. Entries for galaxy pairs and groups, as well as galaxies 
with blueshifts, have been excluded. The data has been supplemented by the recent release 
of 6dF data (Read et al.~2002), but with targets having $cz < 200$ km s$^{-1}$ excluded as they were more likely 
to be galactic. The combined database provides a sample of over 35\,000 galaxies.

Astonishingly, from the core ($cz < 250$ km s$^{-1}$) to the boundary of the sample volume 
($cz = 7500$ km s$^{-1}$), the number density of galaxies declines by a factor of more than 
300 times. In part this is due to the core lying within a large-scale structure, while local 
voids are incorporated in the complete sample volume; this could account for a factor of $\sim$6 times. 
The balance -- a factor of some fifty times -- mainly reflects the increasing incompleteness with 
distance, even though a redshift of $cz = 7500$ km s$^{-1}$ would be considered 
extremely modest compared to modern redshift surveys.

Because of this severe incompleteness problem, our approach here has been to work 
within spherical volumes of increasing radii -- from a radius of $cz = 250$ km s$^{-1}$ out to 
the 7500 km s$^{-1}$ limit -- and to monitor the trends behind measured parameters.

\section{Minimal spanning trees}

{Minimal spanning trees} (hereafter MSTs) allow structures or sub-structures to be 
identified, according to a specified (maximum) percolation distance (e.g. Bhavasar \& Splinter 1996, Krzewina \& Saslaw 1996). They 
form the ideal tool for the current investigation.

The data have been analysed by means of software that progressively creates ever 
larger MSTs. The program, written by one of us (D.T) first passes through the data 
to determine all separations between galaxies. It then works with ever increasing 
{percolation} distance -- starting from zero -- to build up MSTs. Initially numerous 
separate trees form, and grow as the percolation distance increases. In time, as 
the percolation distance increases still further, the trees begin to merge. If run 
to the ultimate conclusion, all trees merge into a single MST.

This scenario is apparent from Figure 1, which shows the number of MSTs within a 
certain data set as a function of  percolation distance. The tabular data from 
which that figure is prepared also records the number of galaxies involved in the tree structures.

One version of the program allows the process to be visualised, so that one can see the 
trees build up and merge (as demonstrated during the oral presentation of this paper). 
That version also allows the programme to be halted at any point. 
A further aspect of the software is that it allows the viewer to fly around the data and 
its MSTs (and can even be viewed in 3D). Figure 2 shows an example of the visualisation.

\begin{figure}[!ht]
\centerline{\hbox{\psfig{figure=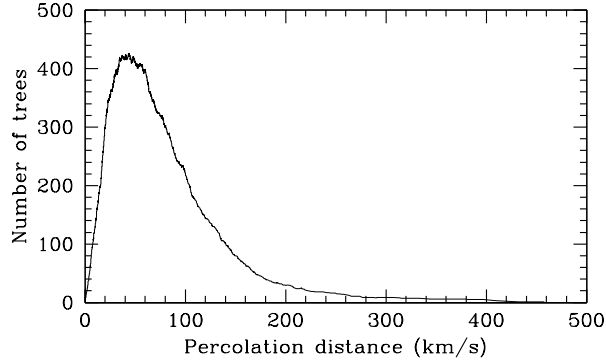,width=8.0cm}}}
  \caption{As the percolation distance increases from zero, so the number of 
MSTs in the data set increases. A maximum is reached, after which trees merge and the 
number declines. Eventually all the data is united as a single tree. 
In this case the data set out to $cz = 1500$ km s$^{-1}$ has been used.}
 \label{fig1}
\end{figure}

\begin{figure*}[!ht]
\centerline{\hbox{\psfig{figure=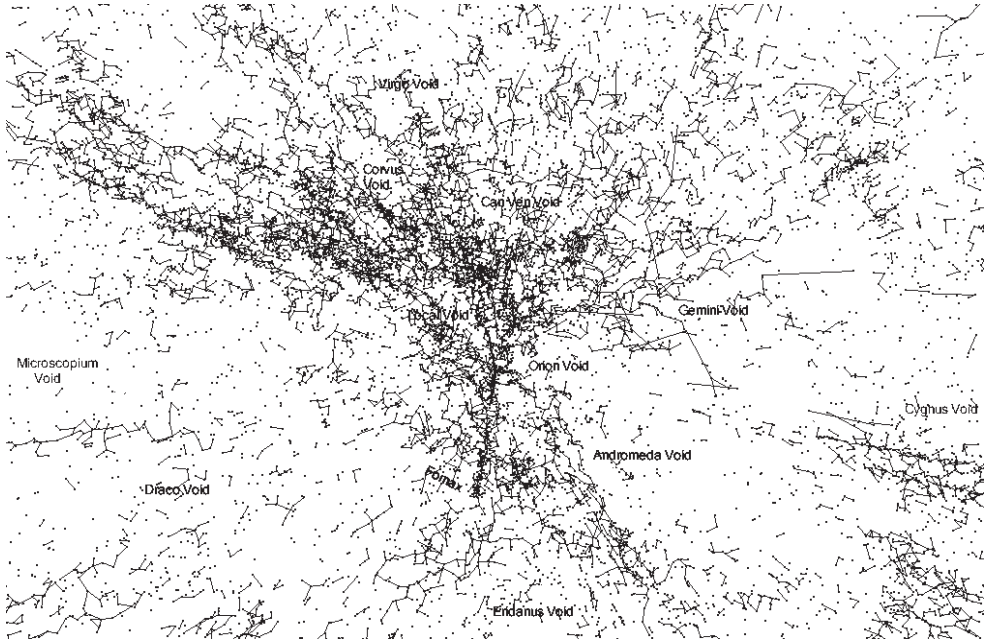,width=13.0cm}}}
  \caption{Minimal spanning trees to 125 km s$^{-1}$ are visible in this plot 
showing the Local Supercluster. The (obviously static) black and white version 
shown here does not allow stereoscopic depth to be perceived, nor the progressive build up of MSTs.  }
 \label{fig2}
\end{figure*}

\begin{table}[!ht]
\caption{Results of perculation analysis for samples of increasing redshift.}\label{tab:fairallt1a}
\begin{minipage}[t]{\textwidth}
\begin{center}
\begin{tabular}{rrrrr}
\hline 
 Data limit & $n$ & Density & Peak & Max \\
(km/s) &   & (10$^{-4}$ Mpc$^{-3}$) & (km/s) & (km/s)\\
\hline
 250  &       242 & 13500 &  11.5  &   97\\
 500  &       492 &  3420 &  20.9  &  249\\
1000  &      1485 &  1292 &  23.2  &  344\\
1500  &      3082 &   794 &  44.1  &  458\\
2000  &      4915 &   534 &  47.9  &  516\\
2500  &      6510 &   362 &  56    &  679\\
3000  &      8510 &   274 &  53.3  &  809\\
4000  &     12375 &   168 &  60.5  &  717\\
5000  &     18334 &   128 &  62.5  & 1014\\
7500  &     35224 &    72 &  77.5  & 1340\\
\hline 
\end{tabular}
\end{center}
\end{minipage}
\end{table}

\begin{table}[!ht]
\caption{Percentage of galaxies included in MSTs.}\label{tab:fairallt1b}
\begin{minipage}[t]{\textwidth}
\begin{center}
\begin{tabular}{lrrrrrrrrrrr}
\hline 
 {\small Data} & {\small To} &{\small To} &{\small To} &{\small To} &{\small To} &{\small To} &{\small To} &{\small To} &{\small To} &{\small To} &{\small To} \\
 {\small limit} & {\small 6.25} &{\small 12.5} &{\small 25} &{\small 50} &{\small 75} &{\small 100} &{\small 125} &{\small 150} &{\small 200} &{\small 250} &{\small 400} \\
 {\scriptsize (km/s)} & {\scriptsize km/s} &{\scriptsize km/s} &{\scriptsize km/s} &{\scriptsize km/s} &{\scriptsize km/s} &{\scriptsize km/s} &{\scriptsize km/s} &{\scriptsize km/s} &{\scriptsize km/s} &{\scriptsize km/s} &{\scriptsize km/s} \\
\hline
 250  &   25.2  & 50.0 &  74.8 & 93.0&  97.9&  100&  100&  100&  100&  100&  100  \\
 500  &   16.5  & 34.6 &  59.8 & 84.1&  92.8& 97.7& 99.2& 99.8& 99.8&  100&  100  \\
1000  &    8.1  & 17.6 &  43.1 & 72.8&  87.2& 92.5& 95.6& 97.6& 99.2& 99.7&  100  \\
1500  &    6.1  & 13.1 &  33.4 & 65.2&  82.3& 90.0& 94.3& 96.6& 98.8& 99.5& 99.9  \\
2000  &    5.2  & 10.9 &  27.8 & 59.6&  78.2& 87.3& 92.5& 95.4& 98.1& 99.1& 99.9  \\
2500  &    4.8  &  9.9 &  25.2 & 55.6&  74.3& 84.2& 90.4& 93.8& 97.3& 98.7& 99.8  \\
3000  &    4.8  &  9.4 &  23.5 & 53.0&  71.5& 81.6& 88.1& 92.1& 96.4& 98.0& 99.7  \\
4000  &    4.1  &  8.1 &  20.1 & 47.0&  65.4& 76.5& 83.7& 88.7& 94.0& 96.7& 99.4  \\
5000  &    3.3  &  6.7 &  17.0 & 41.3&  59.9& 71.9& 80.3& 85.8& 92.2& 95.7& 99.2  \\
7500  &    2.7  &  5.3 &  13.6 & 34.5&  51.8& 64.4& 73.5& 80.1& 88.4& 93.0& 98.2  \\
\hline 
\end{tabular}
\end{center}
\end{minipage}
\end{table}

The outcomes of the MST analyses are shown in Tables 1 and 2. The ten rows in those tables show that the 
procedure described above was run ten times, for ever larger samples, with distance limits (in redshift space) ranging 
from $cz \le 250$ km s$^{-1}$ to $cz \le 7500$ km s$^{-1}$ (the full set of downloaded data), as indicated in the 
first column. The second column  shows the number of galaxies in each of these sample volumes. 
The third column shows the dramatic decline in the density of the galaxies, due to the 
incompleteness of the data (as discussed above).

The position of the peak of the curve (as shown in Figure 1) varies with the sample distance, as 
reflected in the fourth column of Table 1. Had it not been for the incompleteness of the data, the 
position of the peak would have, presumably, not shifted. Note, however, the rapid change in the 
first four data sets, after which the change is more uniform. Coincident with the jump is a change 
in the shape of the actual peak; it is asymmetric for the three sets with $cz < 1000$ km s$^{-1}$, favouring 
the lower values of percolation distance, hence the lower values in this column. For the other 
data sets, the peak is more symmetric (much as it appears in Figure 1) and favours the higher values 
seen in the table.

Table 2 shows what percentages of the galaxies are included in 
MSTs run to certain percolation distances (to 6.25 km s$^{-1}$, to 12.5 km s$^{-1}$ etc).
In spite of the dramatic differences in galaxy density, as one moves (down the table) to 
larger distance limits and sizes of databases, the variations in the percentages are relatively mild.

\begin{figure}
\centerline{\hbox{\psfig{figure=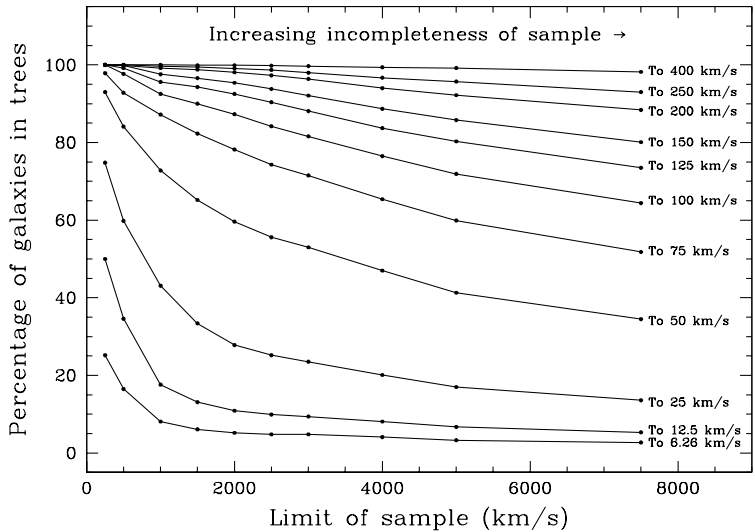,width=10.0cm}}}
  \caption{The percentage of galaxies included in MSTs varies according to both the data set used (horizontal axis) and the maximum allowed percolation distance (separate curves). See text for interpretation.}
 \label{fig3}
\end{figure}

Figure 3 depicts the variations in graphical form. The eleven curves plotted show the 
eleven (maximum) percolations of the MSTs. Their horizontal variations show how they 
decline -- from their true values -- with increasing incompleteness of data. The true values 
ought to be obtained by extrapolating the curves leftward to the zero limit of the sample, but caution needs to be exercised.

First, it can be seen that the lower curves rise steeply upward towards the left of the 
plot. This indicates the presence of a large population of intrinsically faint galaxies, only detectable 
at very low redshifts. It raises the well-known debate about the faint end of the 
luminosity function, with opinions differing as to the quantity of galaxies present. 
Even the lowest of the eleven plots - that for the ridiculously low value of 6.25 km s$^{-1}$ - 
looks as if it might rise dramatically towards the 100\% level. However, in the absence of 
a complete volume limited sample, we cannot be sure that any of the lower five curves 
could or could not reach 100\%. The only curve, for which we have firm evidence of it touching 100\%, 
is the next one up, that for a maximum percolation distance of 100 km s$^{-1}$.

The 
results here strongly suggest that every galaxy has a neighbour within 100 km s$^{-1}$, or 
possibly less, in redshift space.

The final column of Table 1 shows the maximum percolation distance that the trees reach, 
at the point where all trees merge into a single entity; in other words it is the minimum 
percolation distance that would unite all galaxies in the data set. By contrast to the 
percentages, but not unexpectedly, it does show a large variation --  mainly because of 
the increasing incompleteness of the data as the limiting distance is increased. Again 
we can display it in graphical form, this time in Figure 4.

\begin{figure}
\centerline{\hbox{\psfig{figure=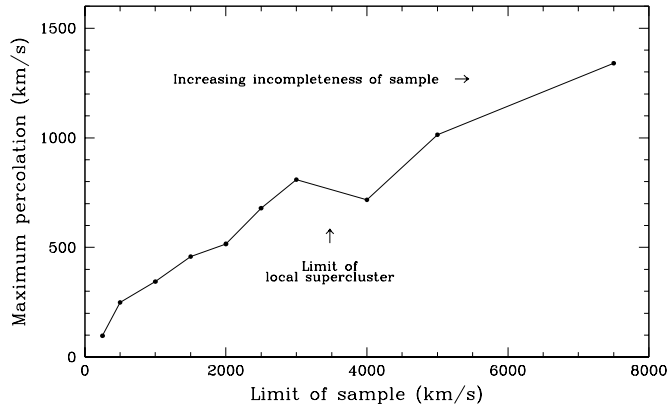,width=8.8cm}}}
  \caption{The value of percolation distance required to link all galaxies into a single MST 
varies with the size of the data sample.}
 \label{fig4}
\end{figure}

To find the true value represented in the figure, we are again faced with an extrapolation 
to the left, and reflecting the experience in the previous figure, the line starts to curve 
as the axis is approached. However, a conservative estimate would be to adopt a value of 100 
km s$^{-1}$, both because it is reached with the smallest data set, and because it would be indicated 
by a linear regression through the seven lowest data points.

It also matches the previous value and confirms that all galaxies have neighbours within 
100 km s$^{-1}$, or possibly less, and, similarly, such a percolation distance serves to interconnect all galaxies, at least 
those within the same large-scale structure. This is an improvement on the working 
definition originally advanced by the first author (F98) for galaxy-galaxy 
separations within a large-scale structure. However, it assumes almost a volume-limited sample.

Figure 4, however, shows some discontinuity beyond the sample size of $cz = 3000$ km s$^{-1}$. 
This is not surprising as this represents the point where the data set expands beyond 
our local large-scale structure, the {Virgo Supercluster}. It raises the question as to what percolation distance 
would unite such large-scale structures. We have only a handful of such superclusters within $cz \le 7500$ 
km s$^{-1}$ (i.e. Virgo, Centaurus, Perseus-Pisces, Fornax, Sculptor and Coma)to go on, but the offset of the line to 700 km s$^{-1}$ suggests that as an upper limit. In short 
we interpret Figure 4 to say that all superclusters (large-scale structures) are interlinked at a 
percolation distance of 700 km s$^{-1}$ (10 Mpc) or less. Thus all large-scale structures interconnect 
to form a single labyrinth of high density regions, as suggested earlier in the `bathsponge' analogy.

These findings are of course based only on data with $cz \le 7500$ km s$^{-1}$, but hopefully - 
though not necessarily - they also describe the character of the nearby universe.

\section{Discussion}

This investigation was prompted by the need to improve upon a definition for large-scale 
structures, yet it seems to reveal something more fundamental about the distribution of galaxies.

Every galaxy has a neighbour (within 100 km s$^{-1}$ = 1.4 $h_{70}^{-1}$ Mpc), and that neighbour 
has a neighbour, such that galaxies are arrayed in {filaments}, or {tree structures} (as seen in Figure 2). This is also 
apparent in Figure 3 of Karachentsev et al. (these proceedings) and Figure 1 of Koribalski 
(these proceedings), both of whom use samples that include low (optical) luminosity galaxies. Karachentsev uses the term the `{Local Cosmic Web}', to describe the filamentary structures, Koribalski `a beautiful network of filaments'. Koribalski also states that `HI surveys also significantly enhance the connectivity of large-scale filaments by filling in gaps populated by gas-rich galaxies'. From the general observation that `slices of the universe' reveal a `bubbly' texture (Koribalski mentions `loops') the 
filaments and tree structures surround voids (much like the filaments in the shells of supernova remnants). On a larger scale 2MASS galaxies (Huchra, these proceedings) portray the `Cosmic web' to a greater depth.

The finding that galaxies are confined to filaments and trees extends the earlier finding that nearby galaxies were predominantly members of groups (see Burstein (Galaxy Groups), these proceedings, discussing Brent Tully's {\em Nearby Galaxy Atlas}). Similarly, in the open discussion at this conference, consensus emerged that there were 
no such things as `field galaxies'. All galaxies are somehow connected to some structure or other.

Clearly this tells us something fundamental about how and where galaxies were formed. However, since their formation 
gravity will have modified their distribution. In the same way that gravity has arrested and 
reversed the cosmological expansion of our {Local Group} (such that the {Great Galaxy in Andromeda} 
and our Galaxy are now drawn towards one another) so the same must hold for similar sized groups. 
This is apparent from the peaking of the histograms (Figure 1) showing how gravity has 
fragmented the distribution into a great number of groups (minimal spanning trees).

The finding is generally in line with the overall scenario that large-scale structures formed 
from the condensation of overdensities; it may, however, suggest that galaxies might have 
formed by some sort of chain reaction. If current n-body simulations (e.g. Pearce et al 2001) accurately 
portray the formation of large-scale structures (if not individual galaxies), then they too 
should exhibit similar percolation properties to those found here. This will provide the basis 
for further investigations.

\section{Background and Acknowledgements}

This work developed from student projects (carried out by co-authors McBride, 
Wiehahn, Pretorius and de Vaux), and was first presented in poster form at IAU Symposium 216 {\em Mapping the Cosmos}, 
associated with the IAU General Assembly in Sydney in July 2003. It has made use 
of the NASA/IPAC Extragalactic Database (NED) which is operated by the Jet Propulsion 
Laboratory, California Institute of Technology, under contract with the National Aeronautics and Space Administration.

\end{document}